\documentclass[english,aps,floats,twocolumn,showpacs,prl,amsfonts]{revtex4}

\usepackage{pslatex}
\usepackage[T1]{fontenc}
\usepackage[latin1]{inputenc}

\usepackage{graphicx}
\usepackage{amssymb}

\makeatletter
\usepackage{babel}
\makeatother

\def\sP{\mathcal{ P}}
\def\sC{\mathcal{C}}
\def\sL{\mathcal{L}}

\def\ssC{\scriptstyle C}

\def\ssM{\scriptstyle M}
\def\ssS{\scriptstyle S}

\def\dsC{\mathcal{C}^\dagger}
\newcommand{\rarrow}{\rightarrow}
\newcommand{\state}[1]{|#1\rangle}

\newcommand{\be}{\begin{equation}}
\newcommand{\ee}{\end{equation}}
\newcommand{\bea}{\begin{eqnarray}}
\newcommand{\eea}{\end{eqnarray}}

\begin{document}

%\begin{frontmatter}

\title{Quantum Mechanical stability of fermion-soliton systems}
\author{Narendra Sahu}
\email{narendra@phy.iitb.ac.in} 
\author{Urjit A. Yajnik}
\email{yajnik@phy.iitb.ac.in}

%\author{Narendra Sahu\ }\ead{narendra@phy.iitb.ac.in}
%and
%\author{Urjit A. Yajnik}\ead{yajnik@phy.iitb.ac.in}

%\address
\affiliation{Department of Physics, Indian Institute of Technology, Bombay,
%\\
Mumbai 400076, India}

%\date{}

\begin{abstract}
Topological objects resulting from symmetry breakdown 
may be either stable or metastable 
depending on the pattern of symmetry breaking. 
However, if they acquire zero-energy modes of fermions, 
and in the process acquire non-integer fermionic 
charge,  the metastable configurations also get stabilized. 
In the case of Dirac fermions the spectrum of the number
operator shifts by $1/2$. In the case of majorana fermions 
it becomes useful to assign negative values  of fermion
number to a finite number of states occupying the 
zero-energy level, constituting a \textit{majorana pond}. 
We determine the parities of these states
and prove a superselection rule. Thus decay of objects
with half-integer fermion number is not possible  
in isolation or by scattering with ordinary particles.
The result has important bearing on cosmology as well as
condensed matter physics.
\end{abstract}

%\begin{keyword}
%Fermion zero-modes \sep Topological objects \sep metastable
%\PACS
%\end{keyword}

%\end{frontmatter}
\maketitle

\section{Introduction}
Solitons present the possibility of extended objects 
as stable states within Quantum Field Theory. Although
these solutions are obtained from semi-classical arguments
in weak coupling limit, their validity as quantal states 
is justified based on the associated topological conservation 
laws. 
A more  curious occurrence is that of fermionic zero-energy modes
trapped on such solutions. Their presence requires,
according to well known arguments\cite{JandR}\cite{Jrev}, an
assignment of half-integer fermion number to the
solitonic states. In the usual treatment, the back reaction 
of the fermion zero-modes on the soliton itself is ignored. 
However, the fractional values of the fermionic charge have 
interesting consequence for the fate of the soliton if 
the latter is not strictly stable.
The reason for this is that if the configuration
were to relax to trivial vacuum in isolation, there is 
no particle-like state available for carrying the fractional 
value of the fermionic charge. Dynamical stability of such objects
was pointed out in \cite{devega}, in cosmological context in \cite{stern},
\cite{steandyaj} and more recently
in \cite{Davis:1997bu}\cite{Davis:1999wq}\cite{DavKibetal}. 
Fractional fermion number phenomenon also occurs in condensed
matter systems and its wide ranging implications call for
a systematic understanding of the phenomenon.

The impossibility of connecting half-integer valued states 
to integer valued states suggests that a superselection 
rule\cite{www} \cite{Sweinberg} is operative. 
In a theory with a conserved charge (global or local), a 
superselection rule operates among sectors of distinct charge 
values because the conservation of charge is associated 
with the inobservability of rescaling operation $\Psi\rightarrow 
e^{iQ}\Psi$. 
For the case of the Dirac fermions, the gauge symmetry is broken, 
however the overall phase
corresponding to the fermion number continues to be a symmetry
of the effective theory. This permits independent rescaling
of the sectors with different values of the fermionic number;
thus superselecting bosonic from fermionic sectors. In the case
at hand, half-integer values occur, preventing such states from 
decaying in isolation to the trivial ground state \cite{devega}
\cite{stern}. 

For the case of majorana fermions where the number operator
is not conserved by interactions the validity of such results
is far from obvious. However, the occurrence of half-integral
values can be shown  to be intimately connected to the charge 
conjugation invariance\cite{sudyaj} of the theory. Here we 
show that in such a case it is possible to assign parities to
the lowest energy solitonic sector which are compatible
with parity assignment in the vacuum sector. A superselection
rule can then be proved for majorana fermions.

%Indeed, the example of \cite{GolWil} shows that arbitrary real
%values of trapped fermion number are possible in a
%theory that does not respect charge conjugation invariance.

In the following we begin with constructing explicit examples 
of cosmic strings which are metastable and have odd number of 
zero-modes, for both the cases of Dirac and  Majorana masses. 
The masses in each case are derived from spontaneous
symmetry breaking. We then discuss assignment of fermion number
to topological objects, including the need for a finite spectrum
of negative values for states of trapped majorana fermions.
Subsequently we derive the required superselection rule. The 
paper ends with summary and conclusion.

\section{Examples}\label{sec:toyexa}
Here we construct two examples in which the topological objects of 
a low energy theory are metastable due to the embedding of the 
low energy  symmetry group in a larger symmetry group at higher 
energy. Examples of this kind were considered in \cite{presvil}.
Borrowing the strategies for bosonic sector from there, we include 
appropriate fermionic content to ensure the zero-modes.

\subsection*{\textbf{A} Dirac fermions}
 
Consider first a model with two stages of symmetry breaking similar
to \cite{presvil}, but with local $SU(2)$ gauge invariance. 
The two scalars $\vec \Sigma$ and $\sigma$ are respectively real 
 triplet and complex doublet.  The Lagrangian is taken to be 
%\cite{presvil}
\begin{eqnarray}
\sL &=& 
-\frac{1}{4} F^{\mu\nu a}F^a_{\mu\nu}
+  \frac{1}{2} D_\mu {\vec \Sigma} \cdot D^\mu \vec\Sigma 
+ D_\mu\sigma^\dag D^\mu\sigma   \nonumber \\
&-&\lambda_1(\vec\Sigma\cdot\vec\Sigma - \eta_1^2)^2
-\lambda_2(\sigma^\dag\sigma - \eta_2^2)^2  \nonumber \\
&+&\lambda_{12}\eta_1 \vec\Sigma\cdot\sigma^\dag{\vec \tau}\sigma
\label{case_one}
\end{eqnarray}
where isovector notation is used and $\tau$ are the Pauli 
matrices. It is assumed that $\eta_1\gg\eta_2$ and that the 
coupling $\lambda_{12}>0$ satisfies $\lambda_{12}\eta_1^2\ll$
$\lambda_2\eta_2^2$

The vacuum expectation value (VEV) 
$\vec\Sigma = (0\  0\ \eta_1)^T$ 
%\qquad  \sigma = \pmatrix{0\cr \eta_2} \ee
breaks the $SU(2)$ to the $U(1)$ generated by $\exp(i\tau^3
\alpha/2)$. The effective theory of the $\sigma$ can be
rewritten as the theory of two complex scalar $\sigma_u$
and $\sigma_d$ for the up  and the down components respectively.
The potential of the effective theory favors the minimum
\be
\sigma_u=\eta_2, \qquad \sigma_d=0
\label{eq:vev}
\ee 
In the effective theory $\sigma_u$ enjoys a $U(1)$ invariance 
$\sigma_u \rarrow e^{i\alpha}\sigma_u$ which is broken by the
above VEV to $\mathbb{Z}\equiv\{e^{2n\pi i}\}$, $n=1,2,\ldots$. 
This makes possible vortex solutions
with an ansatz in the lowest winding number sector
\be
\sigma_u(r, \phi) = \eta_2 f(r) e^{-i\phi}
%; \quad \Sigma_3 = \eta_1
\label{eq:str_ansatz}
\ee
with $r, \phi$ planar coordinates with vortex aligned along the $z$ axis.
The vortex configuration is a local minimum, however it
can decay by spontaneous formation of a monopole-antimonopole
pair \cite{presvil}. These monopoles are permitted by the 
first breaking $SO(3) \rarrow SO(2)$ in the $\Sigma$ sector.
Paraphrasing  the discussion of \cite{presvil}, we have 
$SU(2)\rarrow$$U(1)\rarrow \mathbb{Z}$. 
The vortices are stable in the low energy theory because 
$\pi_1(U(1)/\mathbb{Z})$ is nontrivial. But in the $SU(2)$,
the $\mathbb{Z}$ lifts to $\{e^{4n\pi i\tau^3/2}\}=I$ 
making it possible to unwind the vortex by crossing an energy 
barrier.

Consider now the introduction of a doublet of fermion species
$\psi_L\equiv (N_L, E_L)$ assumed to be left handed and
a singlet right handed species $N_R$. The Yukawa coupling of these
to the $\sigma_u$ is given by $h{\overline{N_R}}\sigma^\dag\psi_L$,
which in the vortex sector reads
\be
\sL_{\sigma-\psi} \sim h \eta_2 f(r)(e^{-i\phi}
{\overline{N_R}} N_L + h.c. )
\label{fermion_case_one}
\ee
The lowest energy bound states resulting from this coupling
are characterized by a topological index, \cite{Eweinberg}
\( \mathcal{I} \equiv n_L - n_R\) where 
$n_L$ and $n_R$ are the zero modes of the left handed and 
the right handed fermions respectively. This index can be computed
using the formula \cite{Eweinberg}\cite{GanLaz}
\be
\mathcal{I} = \frac{1}{2\pi i}(\ln \textrm{det}M)\vert^{2\pi}_{\phi=0}
\label{eq:index}
\ee
where  $M$ is the position 
dependent effective mass matrix for the fermions. 
In the present case this gives rise to a single zero-energy mode 
for the  fermions of species $N$. According to well known reasoning 
\cite{JandR} to be recapitulated 
%in sec. \ref{sec:ferno} 
below, this  requires the assignment of either of the values 
$\pm1/2$ to the fermion number of this configuration. 

\subsection*{\textbf{B} Majorana fermions}
The example above can be extended to the case where the $N$ is a
majorana fermion. Being a singlet $N$ admits a mass term
$M_{\ssM} {\overline{N_R^{\ssC}}}N_R$, $M_{\ssM}$ signifying majorana mass. 
This could also be a spontaneously 
generated mass due to the presence of a
neutral scalar $\chi$ with coupling terms 
$h_{\ssM}{\overline{N_R^{\ssC}}}N_R\chi + h.c.$. 
If this $\chi$ acquires a VEV at energies higher than the 
$\Sigma$, the $N$ particles possess a majorana mass and
fermion number is not a conserved observable.

Finally we present the case where majorana mass is spontaneously
generated at the same scale at which the vortex forms. 
Consider a  theory  with local $SU(3)$ symmetry broken 
to $U(1)$ by two scalars, $\Phi$ an octet acquiring a VEV $\eta_1\lambda_3$ 
($\lambda_3$ here being the third Gell-Mann matrix) and 
$\phi$, a ${\bar 3}$, acquiring the VEV $\langle\phi^k\rangle=\eta_2
\delta^{k 2}$, with $\eta_2\ll \eta_1$. Thus
\be
SU(3) {\buildrel 8 \over{\longrightarrow}}
U(1)_3\otimes U(1)_8 { \buildrel \overline{3}  \over{\longrightarrow}} U(1)_+
\ee
Here $U(1)_3$ and $U(1)_8$ are  generated by $\lambda_3$ 
and $\lambda_8$ respectively, and $U(1)_+$ is generated by
\( (\sqrt{3}\lambda_8 + \lambda_3)/2 \) and likewise $U(1)_-$
to be used below.  
It can be checked that this pattern of VEVs can be generically
obtained from the quartic scalar potential of the above Higgses. 
The effective theory at the second breaking $U(1)_-\rightarrow \mathbb{Z}$ 
gives rise to cosmic strings. However the $\mathbb{Z}$ lifts to identity in 
the $SU(3)$ so that the string can break with the formation of 
monopole-antimonopole pair.

Now add a multiplet of left-handed fermions belonging to $\overline{15}$. 
Its mass terms arise from the following coupling to the $\overline{3}$
\be
\sL_{\textrm{Majorana}} = h_M \overline{\psi^{\ssC}}^{\{ij\}}_{k}
\psi^{\{lm\}}_{n}\phi^{r}
(\epsilon_{ilr}\delta^{n}_{j}\delta^{k}_{m})
\ee
The indices symmetric under exchange have been indicated by curly brackets. 
No mass terms result from the $8$ because it cannot provide a singlet
from tensor product with $\overline{15}\otimes \overline{15}$ 
\cite{Slansky:yr}. 
After substituting the $\phi$ VEV a systematic 
enumeration shows that all but the two components $\psi^{\{22\}}_{1}$ 
and $\psi^{\{22\}}_{3}$ acquire
majorana masses at the second stage of the breaking. Specifically 
we find the majorana mass matrix to be indeed rank $13$. Thus, using 
either of  the results \cite{JandRossi} or \cite{GanLaz} i.e.,\  eq. 
(\ref{eq:index})
we can see that there will be $13$   zero modes present in the lowest 
winding sector of the cosmic string. Thus the induced fermion number 
differs from that of the vacuum by half-integer as required.

\subsection*{\textbf{C} Final state zero-modes}
\begin{figure}[htb]
{\par\centering \resizebox*{0.3\textwidth}{!}
{\rotatebox{0}{\includegraphics{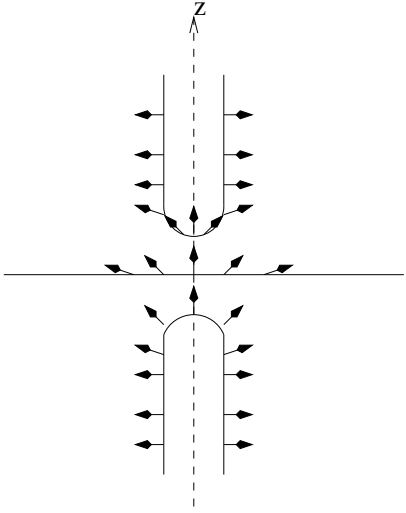}}} \par}\vspace{0.3cm} 
% (string.pdf for pdfLaTeX, string.ps for pslaTeX or LaTeX)
\caption{Schematic configuration of isospin vectors after the
rupture of a string. Internal orientations are mapped to external 
space. They are shown just outside the core of the two resulting 
pieces and on the  mid-plane symmetrically separating the two.}
\label{fig:string}
\end{figure}

The stability argument being advanced is in jeopardy if the final
state after rupture of the topological object also possesses half-integral
fermionic charge.  To see that this is not the case it is necessary to study
the zero-modes on the two semi-infinite strings shown
in fig. \ref{fig:string}. Generically we expect each of the halves
to support the same number of zero-modes, making the total 
fermion number of the putative final state integer valued, as required
for the validity of our argument.

Consider the ansatz for the lower piece ($l$) with origin at the
corresponding monopole and coordinates \((r_l, \theta_l, \phi)\).
For the domain $z<0$ let the ansatz for the field $\sigma$ be
\be
U_{l}^\infty(\theta_l,\phi) \pmatrix{0 \cr \eta_2} f_l(r_l)
\equiv  \exp\{\frac{i}{2}\theta_l {\vec \tau}\cdot \hat{\phi} \}
\pmatrix{0 \cr \eta_2} f_l(r_l)
\ee
so that \( \langle \sigma \rangle \) has the behaviour
\begin{displaymath}
\langle\sigma\rangle = \left\{ \begin{array}{lr}
\pmatrix{0 \cr \eta_2}f_l(r_l) & \textrm{for}\ \theta_l\approx 0\\
& \\
\pmatrix{\eta_2e^{-i\phi} \cr 0}f_l(r_l)  & \textrm{for}\ \theta_l\approx \pi
\end{array}
\right.
\end{displaymath}
which agrees with the ansatz (\ref{eq:str_ansatz})  for the cosmic string
at the South pole. Likewise for the domain $z>0$, ie the upper piece ($u$), 
we choose 
\be
U_{u}^\infty(\theta_u,\phi) = \exp\{\frac{i}{2}(\pi-\theta_u) 
{\vec \tau}\cdot \hat{\phi} \}
\ee
resulting in  the behaviour
\begin{displaymath}
\langle\sigma\rangle = \left\{ \begin{array}{lr}
\pmatrix{\eta_2e^{-i\phi} \cr 0}f_u(r_u) & \textrm{for}\  \theta_u\approx 0\\
 &  \\
\pmatrix{0 \cr \eta_2}f_u(r_u) & \textrm{for}\  \theta_u\approx \pi
\end{array}
\right.
\end{displaymath}
thus matching correctly with the cosmic string at the North pole.
The ansatz for the heavier scalar $\Sigma$ needs to be appropirately set up, 
\( U^\infty {\vec\Sigma} \cdot{\vec\tau} {U^\infty}^\dag \)
in both $l$ and $u$ domains. This scalar however does not contribute to 
fermion mass matrix. 

The two maps match at the mid-plane where $\theta_l=\pi - \theta_u$
and \( \sigma_u \sim e^{-i\phi} \) at \( \theta_l=\theta_u =\pi/2\)
so that we have ensured that the combined map is within the same
homotopy class as the string we began with. Finally, as the two 
pieces move far away, each can be seen to have the same number of
zero-modes. To see this we can choose \cite{DHN}\cite{nohl} fermion
ansatz for the zero-modes compatible with the scalar field ansatz, 
in each of the patches $l$ and $u$. 
In (isospin)$\otimes$(two component spinor)
notation for $\psi_L$ and for the two component fermion $N_R$,
\be
\psi_L = U^\infty(\theta,\phi) \pmatrix{0 \cr 1}\otimes
\pmatrix{\varphi_1(r) \cr \chi_1(r)}
\quad
N_R = \pmatrix{\varphi_2(r) \cr \chi_2(r)}
\ee
where the labels $l$, $u$ have been dropped. To analyse the asymptotic
radial dependence choose $\gamma^{\ r}=\sigma^2$ the Pauli matrix.
In each patch 
one finds the pair \( \varphi_1(r), \chi_2(r) \sim e^{-h\eta_2 r}\)
to constitute the zero-mode while for the other pair, \( \varphi_2(r), \chi_1(r) 
\sim e^{+h\eta_2 r}\) which are therefore not normalizable.
In any case, since each of the pieces acquires  the same number 
of zero-modes, the total fermion number of the putative
final state has been proved to be integer as required.

\section{Assignment of fermion number}\label{sec:ferno}
We now recapitulate the reasoning behind the assignment of
fractional fermion number. We focus on the Majorana fermion
case, which is more nettlesome, while the treatment of the
Dirac case is standard \cite{JandR}\cite{Jrev}. In the prime example 
in $3+1$ dimensions 
of a single left-handed fermion species $\Psi_L$ coupled to an abelian
Higgs model according to 
\be
\mathcal{L}_\psi = i\overline{\Psi_L}\gamma^\mu D_\mu \Psi_L
-\frac{1}{2}( h \phi \overline{\Psi_L^C}\Psi_L + h.c.)
\label{fermion_case_two}
\ee
the following result has been obtained\cite{JandRossi}.
For a vortex oriented along 
the $z$-axis, and in the winding number sector $n$, 
the fermion zero-modes are of the form
\be
\psi_0({\bf x})=\pmatrix{1\cr 0}\left[U(r)e^{il\phi}
+V^*(r)e^{i(n-1-l)\phi}\right]g_l(z+t)
\ee
In the presence of the vortex, $\tau^3$ (here representing
Lorentz transformations on spinors) acts as the
matrix which exchanges solutions of positive frequency
with those of negative frequency. It is therefore identified
as the ``particle conjugation'' operator.
In the above ansatz, the $\psi$ in the zero-frequency 
sector are charge self-conjugates, $\tau^3\psi=\psi$,
and have an associated left moving zero mode along the
vortex. The functions satisfying $\tau^3\psi=-\psi$
are not normalizable. The situation is reversed 
when the winding sense of the scalar field is reversed, ie, for
$\sigma_u\sim$$e^{-in\phi}$. In the winding number sector $n$, 
regular normalizable solutions\cite{JandRossi} exist for
for $0\leq l\leq n-1$. The lowest energy sector of the vortex is 
now \(2^n\)-fold degenerate, and each zero-energy mode needs to be 
interpreted as contributing a value $\pm 1/2$ to the total 
fermion number of the individual states\cite{JandR}. This conclusion
is difficult to circumvent if the particle spectrum is to reflect
the charge conjugation symmetry of the theory \cite{sudyaj}. 
The lowest possible value of the induced number in this sector
is $-n/2$. Any general state of the system is built from one
of these states by additional integer number of fermions.
All the states in the system therefore possess half-integral
values for the fermion number if $n$ is odd. 

One puzzle immediately arises, what is the meaning of
negative values for the fermion number operator for
\textit{Majorana} fermions? In the trivial vacuum, we can
identify the Majorana  basis as
\be
\psi\ =\ \frac{1}{2}(\Psi_L +  \Psi_L^C)
\label{majdef}
\ee
This leads to the Majorana condition which results in
identification of particles with anti-particles according to
\be
\sC \psi \dsC\  =\   \psi 
\label{majcond}
\ee
making negative values for the number meaningless.
Here $\sC$ is the charge conjugation operator.  We shall 
first verify that in the zero-mode 
sector we must indeed assign negative values to the number 
operator. It is sufficient to treat 
the case of a single zero-mode, which generalizes easily
to any larger number of zero-modes.
The number operator possesses the properties
 \be
[ N, \psi ]\ =\ -\psi\qquad{\rm and} 
\qquad [ N, \psi^\dagger ]\ =\ \psi^\dagger
\label{psiN}
\ee
\be
\sC N \dsC\ =\ N
\label{ccN}
\ee
Had it been the Dirac case, there should be a minus sign 
on the right hand side of eq. (\ref{ccN}). This is absent due 
to the Majorana condition. The fermion field operator for the
lowest winding sector is now expanded as
\be
\psi\ =\ c\psi_0 + \left\{ \sum_{{\bf \kappa},s} 
a_{{\bf \kappa},s}\chi_{{\bf \kappa},s}(x)\ \\
+\ \sum_{{\bf k},s} b_{{\bf k},s}u_{{\bf k},s}(x)\ 
+\ h.c. \right\}
\label{eq:expansion}
\ee
where the first summation is over all the possible bound states of 
non-zero frequency with real space-dependence of the form $\sim
e^{-{\bf \kappa\cdot x_\perp}}$ in the transverse space directions
${\bf x}_\perp$, and the second summation is over all 
unbound states, which are asymptotically plane waves. 
These summations are suggestive and their exact connection to
the Weyl basis mode functions \cite{FukSuz}  are not essential 
for the present purpose.
Note however that no "$h.c.$" is needed for the zero energy mode
which is self-conjugate.  Then the Majorana condition (\ref{majcond}) 
requires that we demand
\be
\sC\ c\ \dsC\ =\   c \qquad{\rm and} 
\qquad \sC\  c^\dagger\  \dsC\ =\  c^\dagger
\label{ccc}
\ee
Unlike the Dirac case, the $c$ and $c^\dagger$ are not
exchanged under charge conjugation. 
%and consistency with  standard anti-commutation relations 
%for the $c$ and $c^\dagger$ is retained.
The only non-trivial irreducible realization of this
algebra is to require the existence of a doubly degenerate
ground state with states $\state{-}$ and $\state{+}$ satisfying
\be
c\state{-}\ =\ \state{+}\qquad {\rm and}\qquad 
c^\dagger\state{+}\ =\ \state{-}
\label{cstates}
\ee
with the simplest choice of phases. Now we find
\begin{eqnarray}
\sC\  c\  \dsC \sC \state{-}\ &=\ \sC\state{+}\\ \vspace{3mm}
\Rightarrow \hspace{2mm}  c (\sC \state{-})\ &=\ (\sC\state{+})
\label{Ctransform}
\end{eqnarray}
This relation has the simplest non-trivial solution
\be
\sC\state{-}\ =\ \eta^-_{\ssC} \state{-}\qquad
{\rm and}\qquad \sC\state{+}\ =\ \eta^+_{\ssC} \state{+}
\label{Cproperty}
\ee
where, for the consistency of  (\ref{cstates}) and (\ref{Ctransform})
$\eta^-_{\ssC}$ and $\eta^+_{\ssC}$ must satisfy
\be
(\eta^-_{\ssC})^{-1}\eta^+_{\ssC}\ =\ 1
\ee
Finally we verify that we indeed get values $\pm 1/2$ for $N$.
The standard fermion number operator which in the Weyl basis is
\be
N_F = \frac{1}{2}[ \Psi_L^\dag \Psi_L - \Psi_L \Psi_L^\dag ]
\ee
%\be
%N_{\mathrm{vac}}\ =\ {1\over 2}(\psi^\dagger \psi\ -\ \psi\psi^\dagger) 
%\label{Noperator} 
%\ee
acting on these two states gives,
\be
{1\over 2}(c\ c^\dagger\  -\  c^\dagger\  c)\ \state{\pm}\ =\ 
\pm{1 \over 2}\state{\pm}
\ee
The number operator indeed lifts the degeneracy of the
two states. For $s$ number of zero modes, the ground
state becomes $2^s$-fold degenerate, and the fermion number
takes values in integer steps ranging from $-s/2$ to $+s/2$.
For $s$ odd the values are therefore half-integral.
%As we shall see below, additionally, separate
%parities must be assigned to the two possible states,
%thus identifying physically observable states unambiguously.
Although uncanny, these conclusions accord with some known
facts. They can be understood as spontaneous symmetry breaking 
for fermions\cite{mcwilczek}. The negative values of the number thus
implied occur only in the zero-energy sector and do not 
continue indefinitely to $-\infty$. Instead of an unfathomable 
\textit{Dirac sea} we have a small \textit{Majorana pond} 
at the threshold.

\section{Quantum Mechancial stability}\label{sec:indsta}

The theory of eq. (\ref{case_one}) possesses a gauge symmetry which
is  reflected in the effective theory (\ref{fermion_case_one}) as 
$N_L\rarrow$$ e^{i \alpha} N_L$, $N_R\rarrow$$ e^{i \alpha} N_R$
giving rise to the usual conserved number for Dirac fermions.
 The lowest winding vortex sector
results in half-integer values for this number. Quantum Mechanical 
stability of this sector follows from well known arguments 
\cite{www}\cite{Sweinberg} which can now be understood
as either following from distinctness of sectors of different values of
$(-1)^{N_F}$, or as a consequence of a residual subgroup of the 
gauge symmetry. For the Majorana case we shall now carry out this 
kind of argument explicitly.

It is known that Majorana fermions can be assigned a unique 
parity \cite{Sweinberg},  either of the values $\pm i$.
Accordingly let us choose $i$ to be the parity of the free single fermion
states in the trivial vacuum.
%%%%%%%%%%%%%%%%%%%%%%
%%%%%%%%%%%%%%%%%%%%%%
As a step towards deriving our superselection rule,
we determine the parities of the zero-energy states. 
The fermion spectrum should
look the same as trivial vacuum far away from the vortex 
\footnote{This should be understood to be the clustering 
property of local field theory being obeyed by the vortex
ground state.}. In turn the parities of the latter states should
be taken to be the same as those of the trivial ground state.
Next, any of these asymptotic free fermions is capable of
being absorbed by the vortex (see for instance \cite{Davis:1999ec}). 
In the zero energy sector this absorption would cause a 
transition from \( \state{-1/2} \) to \( \state{1/2} \)
and cause a change in parity by $i$.
Thus the level carrying fermion number \(+1/2\) should be
assigned a parity $e^{i\pi/2}$ relative to the \(-1/2\) state. 
Symmetry between the two states suggests that we assign parity 
$e^{i\pi/4}$ to the $N_F=1/2$ and $e^{-i\pi/4}$ to the $N_F=-1/2$ 
states. 

Similar reasoning applies to a residual discrete symmetry belonging
to the original $U(1)$ gauge group of Lagrangian (\ref{fermion_case_two}).
According to eq. (\ref{majdef}), under  gauge transformation, 
\be
\psi \rarrow \psi_{[\alpha]} \equiv 
\frac{1}{2}(e^{i\alpha}\Psi_L + e^{-i\alpha}\Psi_L^C)
\ee
Thus $\alpha=\pi$ preserves the choice of the Majorana  basis 
upto a sign.
After  symmetry breakdown and  Higgs mechanism, the 
Yukawa coupling takes the form
$ \sim (m+{\tilde \phi})\overline{\psi} \psi$, which is invariant
under the residual $\mathbb{Z}_2$ symmetry $\psi$  $\rarrow -\psi$. We can use
this as a discrete symmetry distinguishing states of even and odd
majorana fermions. Since single majorana fermions can
be absorbed by the vortex \cite{Davis:1999ec},
the ground states $|\pm \rangle$ are distinguished from each
other  by a relative negative sign. To be symmetric we
can assign the value $\pm i$ to these states under this discrete symmetry 
with sign same as in the value of the number operator. It is
possible to prove the superselection rule using this conserved
quantity. However we also see that this discrete symmetry 
can be used to change our convention of the parity for free 
majorana particles from $+i$ to $-i$. Thus the two are intimately
related and in what follows we shall use the parity with convention
as in the preceding paragraph.

We now show the inappropriateness of superposing states of
half-integer valued fermion number and integer valued fermion
number \cite{www}. The operation $\sP^{\displaystyle4}$, parity 
transformation performed four times  must return the system 
to the original  state, upto a phase. Consider forming the 
state $\Psi_{\ssS}=$$\frac{1}{\sqrt{2}}(\state{1/2}+\state{1})$ 
from states of half-integer and integer value for the fermion number. 
But 
\be
\sP^4\ \Psi_{\ssS}\ =\ {1\over \sqrt{2}}(-\state{1/2}+\state{1})\ 
%\equiv\ \psi_{\ssA}
\ee
Thus this operation identifies a state with another orthogonal
to it. Similarly, application of $\sP^{\displaystyle2}$ which 
should also leave the physical content of a state unchanged
results in yet another linearly independent state, 
$\frac{1}{\sqrt{2}}(i\state{1/2}-\state{1})$.
Thus the space of superposed states collapses to a trivial
vector space. The conclusion therefore is that it is not
possible to superpose such sectors. In turn there can be
no meaningful operator  possessing non-trivial matrix
elements between the two spaces. This completes our proof
of the theorem.

\section{Conclusion}

Metastable classical lumps, also referred to as embedded defects
can be found in several theories.
The conditions on the geometry of the vacuum manifold
that give rise to such defects were spelt out in \cite{presvil}.
We have studied the related question of fermion zero-energy 
modes on such objects. It is possible to construct examples
of cosmic strings in which the presence of zero-modes signals 
a fractional fermion number both for Dirac
and Majorana masses. 
It then follows that such a cosmic string cannot decay in
isolation because it belongs to a distinct superselected 
Quantum Mechanical sector. Thus a potentially metastable object 
can enjoy induced stability due to its bound state with fermions.

Although decay is not permitted in isolation, it certainly
becomes possible when more than one such objects come
together in appropriate numbers. In the early Universe such 
objects could have formed depending on the unifying group 
and its breaking pattern. Their disappearance would be
slow because it can only proceed through encounters between
objects with complementary fermion numbers adding up to an 
integer. Another mode of decay is permitted by change in
the ground state in the course of a phase transition.
When additional Higgs fields acquire a vacuum expectation 
value, in turn altering the boundary conditions for the
Dirac or Majorana equation, the number of induced zero
modes may change from being odd to even\footnote{Such a possibility,
though for topologically stable string can be found for realistic
unification models in \cite{steandyaj} \cite{Davis:1997bu} 
\cite{Davis:1999ec} \cite{widyajetal}.}, thus imparting
the strings an integer fermion number. The decay can then proceed
at the rates calculated in \cite{presvil}. 

In condensed matter systems the best known example is of
charge carriers with half the electronic charge in
polyacetylene \cite{Jackiw:wc} where the underlying
solitons have been identified as stable kinks. It would be 
interesting to find intrinsically metastable objects stabilized
by this mechanism.

Finally it is interesting to conjecture about the fate of 
loops of cosmic strings. The zero modes on closed string loops 
have not been studied in detail, though dynamical stability due 
to superconductivity has been identified. If loops carry 
fractional fermion number and get stabilized by this mechanism 
that also would present very interesting possibilities for Cosmology.

\section{Acknowledgement}
%\begin{acknowledgments}
We thank Sumit Das, P. Ramadevi and S. Uma Sankar 
for their for critical comments on the manuscript.
We are also indebted to the referee for critical comments.
%\end{acknowledgments}

%\bigskip
%\hrule

%\section*{References}

\end{document}